# Navigating the New Landscape: A Conceptual Model for Project-Based Assessment (PBA) in the Age of GenAI

Rajan Kadel, Samar Shailendra, Urvashi Rahul Saxena
*School of IT and Engineering (SITE)*
*Melbourne Institute of Technology*
Melbourne, Australia
rkadel@mit.edu.au, sshailendra@mit.edu.au, usaxena@mit.edu.au

*Abstract*—The rapid integration of Generative Artificial Intelligence (GenAI) into higher education presents both opportunities and challenges for assessment design, particularly within Project-Based Assessment (PBA) contexts. Traditional assessment methods often emphasise the final product in the PBA, which can now be significantly influenced or created by GenAI tools, raising concerns regarding product authenticity, academic integrity, and learning validation. This paper advocates for a reimagined assessment model for Project-Based Learning (PBL) or a capstone project that prioritises process-oriented evaluation, multi-modal and multifaceted assessment design, and ethical engagement with GenAI to enable higher-order thinking. The model also emphasises the use of (GenAI-assisted) personalised feedback by a supervisor as an observance of the learning process during the project lifecycle. A use case scenario is provided to illustrate the application of the model in a capstone project setting. The paper concludes with recommendations for educators and curriculum designers to ensure that assessment practices remain robust, learner-centric, and integrity-driven in the evolving landscape of GenAI.

*Index Terms*—Academic integrity, assessment design, authentic assessment, capstone project, Generative Artificial Intelligence (GenAI), higher education, Project-Based Assessment (PBA), Project-Based Learning (PBL), reflective practice.

## I. Introduction

The rapid advancement of Generative Artificial Intelligence (GenAI) has begun to reshape higher education by influencing how students' knowledge is accessed. GenAI tools are becoming increasingly integrated into students' academic practices, from content generation to idea development and language refinement. This technological shift presents new possibilities for learning and academic support, while also challenging traditional norms of authorship, originality, and skill acquisition.

GenAI poses significant challenges to traditional assessment methods by blurring the lines between student-generated work and GenAI-assisted output. Traditional assessments that emphasise final products may no longer effectively reflect a student's understanding or effort. In Project-Based Assessment (PBA), where collaboration, critical thinking, and iterative development are the key, GenAI can either enhance or compromise the authenticity of the process, depending on how it is used and evaluated.

Given the impact of GenAI, there is a growing need to revisit the design of PBA to ensure they remain valid and educationally meaningful. This involves shifting focus toward the learning process, ethical use of GenAI, and the development of higher-order thinking skills. Redesigning PBA to incorporate transparent GenAI use, process-based evaluation, and multi-modal evidence of learning will help maintain academic integrity while preparing graduates for a future shaped by GenAI.

According to the report "*Assessment reform for the age of artificial intelligence*" by Tertiary Education Quality and Standards Agency (TEQSA), Australia [1], the two principles to be ensured during assessment design in the era of GenAI are:

- Students are expected to engage with GenAI tools in an ethical and informed manner, demonstrating awareness of their limitations, potential biases, and broader implications. Assessment design should account for both the opportunities and risks associated with the use of GenAI, and
- Assessment strategies should adopt diverse, inclusive, and contextually relevant approaches to support meaningful student learning outcomes.

To uphold the outlined principles, academics need to design assessments that not only facilitate student learning of the subject matter but also promote the effective and responsible use of GenAI tools through well-considered assessment strategies.

The primary contributions of this paper are listed below
- Introduce principles for redesigning PBA in the era of GenAI to ensure academic integrity, and learning validation, and
- Proposed an execution model incorporating the proposed design principles and evaluation strategies for PBA in the Era of GenAI.

The outline of the paper is as follows: Section II provides opportunities and challenges of PBA in the GenAI Era. Section III provides a brief review of assessment design, including PBA in the era of GenAI. The proposed PBA assessment redesign principles for PBA are presented in Section IV. The proposed assessment model and use case example with analysis are presented in Section V and Section VI, respectively. Finally, Section VII presents the concluding remarks and the future directions.

## II. Opportunities and Challenges

In this section, the opportunities and challenges associated with GenAI-specific PBA are highlighted. Chiu [2] conducted a scoping review to investigate how GenAI transforms assessment in higher education, identifying both opportunities and challenges. The review emphasises the need for institutions to adapt assessment practices to maintain academic integrity and learning outcomes.

GenAI presents key opportunities to enhance learning by serving as a versatile and dynamic support tool throughout the project-based work [3]. GenAI (as a potential aid) can assist students in brainstorming ideas, generating content, and refining project direction, enabling more efficient planning and clearer articulation of goals. Additionally, it can be leveraged to provide immediate feedback on the content, improve language and structure, and facilitate research, making it a powerful tool for enhancing the quality of work. Furthermore, GenAI also acts as an effective organisational tool, helping students manage tasks, timelines, and resources, which is particularly valuable in project-based work to be completed in group settings. Additionally, its capabilities as a co-creation tool allow students to iterate on their work, experiment with multiple solutions, and engage in creative collaboration. Through structured and responsible use of GenAI in PBA, it fosters the development of critical thinking and professional skills, ultimately contributing to the preparation of industry-ready graduates for AI-augmented work environments.

While GenAI offers valuable assistance in Project-Based Learning (PBL), it also introduces significant challenges that must be carefully addressed in educational settings. A major concern is the risk that excessive reliance on GenAI-generated content may undermine students' learning, leading to superficial understanding rather than meaningful knowledge construction. This raises academic integrity concerns, as traditional detection tools are often insufficient for identifying AI-assisted work, making it challenging to ensure authenticity. Additionally, assessing core competencies such as creativity and critical problem-solving becomes more complex as GenAI can substantially contribute to idea generation and product development. The increasing sophistication of these tools also poses threats to assessment security, calling for redesigned evaluation frameworks that can effectively distinguish student effort from GenAI output and preserve the credibility of educational outcomes. Therefore, there is a need for a proper assessment design to use GenAI tools and technologies in PBL [3]–[5].

## III. Related Works

The advent of GenAI has significantly transformed the landscape of higher education, particularly in the realm of PBA. Traditional assessment methods, which often emphasise the final product, are increasingly challenged by GenAI's capabilities to generate content, raising concerns about authenticity, academic integrity, and the validation of student learning.

Smith and Francis [6] advocate for a shift from product-oriented to process-oriented assessments, emphasising the importance of evaluating the student's journey, including their interactions with GenAI tools. This approach aligns with the need to ensure that assessments capture the depth of student engagement and learning, rather than solely the end product. Divjak et al. [5] explore the intersection of problem-based learning and Artificial Intelligence (AI), suggesting innovative assessment strategies that incorporate AI tools. Their work underscores the potential of AI to enhance learning experiences but also highlights the challenges in maintaining assessment authenticity. Gonsalves [7] discusses the contextual assessment design in the age of GenAI, highlighting the necessity for assessments that account for the diverse ways students interact with AI tools. This perspective supports the development of assessments that are both authentic and adaptable. Belkina et al. [8] conducted a systematic review on implementing GenAI in higher education, noting a predominant focus on general overviews with a shortage of research in specific topics, including integration with teaching practices and AI in assessment. In [9], the authors classify the assessment types in the era of GenAI and outline key steps used in the authentic assessment design. Additionally, GenAI tools are utilised for the content development [10]

Pelleti et al. [11] provide a framework for institutions to evaluate their preparedness for strategic AI initiatives, emphasising the importance of cross-functional collaboration in adapting to AI-driven changes in education at the EDUCAUSE Higher Education AI Readiness Assessment. Shailendra et al. [12] propose frameworks for integrating GenAI into educational assessments, highlighting the necessity for ethical guidelines and the development of AI literacy among students. However, these frameworks often lack detailed strategies for implementing such principles within PBA contexts. The research illustrated in [13] introduces a method for assessing the impact of GenAI on curriculum design, emphasising the need for adaptability in learning outcomes and assessment formats. While this provides a macro-level perspective, there remains a gap in micro-level application within specific assessment tasks.

Bewersdorff et al. [14] conducted a multinational assessment of AI literacy among university students, highlighting the relevance of affective orientations towards AI, such as self-efficacy and interest, in the learning process. Gu and Ericson [15] provide an integrative review of AI literacy in K-12 and higher education, emphasising the need for a unified framework to promote AI capabilities among students. LaFlamme [16] proposes an instructional model for scaffolding AI literacy, addressing the increasing demand for AI literacy in higher education and the importance of preparing students for an AI-integrated world. Sol et al. [17] suggest redesigning learning assessments by adopting or adapting the five scales of AI assessment, promoting inclusive and ethical application of GenAI in higher education while upholding academic honesty.

Several international and national agencies highlight the key focus points on the assessment design in the era of GenAI. The UNESCO report [18] discusses the future of assessment in the AI age, emphasising the need to move beyond traditional metrics that prioritise memorisation and formulaic responses, advocating for assessments that promote deeper learning. The EDUCAUSE report on defining AI literacy for higher education [19] outlines the technical, evaluative, practical, and ethical dimensions of AI literacy, providing a comprehensive framework for institutions. The Digital Education Council's Global AI Faculty Survey [20] reveals a disconnect between high AI usage and low AI competency among both students and faculty, underscoring the urgent need for AI literacy initiatives in higher education. The Australian Council for Educational Research [21] highlights the impact of GenAI on assessment and learning outcomes, emphasising the need for educators to adapt assessment practices to maintain academic integrity.

Elize discusses embracing PBAs in the age of AI, highlighting the benefits of GenAI in enhancing student cooperation and engagement [22]. William et al. [23] explore the future of AI in PBL through a co-design approach with students, emphasising the importance of incorporating student perspectives in developing AI-integrated assessments. The paper discusses the use of AI assistants in business education, highlighting the potential of AI to enhance personalised learning and support educators in grading and assessment tasks [24].

Despite these contributions, there is a lack of a holistic approach to PBA design in the era of GenAI. The key research gaps found from the brief literature review are:

- A few literatures are available in PBA design after the introduction of GenAI,
- Literature tries to address individual concerns rather than a holistic approach, and therefore, there is a need for a holistic approach that addresses process-oriented evaluation; implementation of multi-modal and multi-faceted assessment; cultivating AI literacy as a core learning outcome, including encompassing ethical use, transparency, and bias mitigation; systematic mechanisms to assess higher-order thinking skills, and
- Minimal exploration of personalised feedback loops enhanced by GenAI in the assessment design.

Next, we introduce principles for redesigning PBA considering the introduction of GenAI in an educational landscape, which is followed by an assessment model to address these gaps.

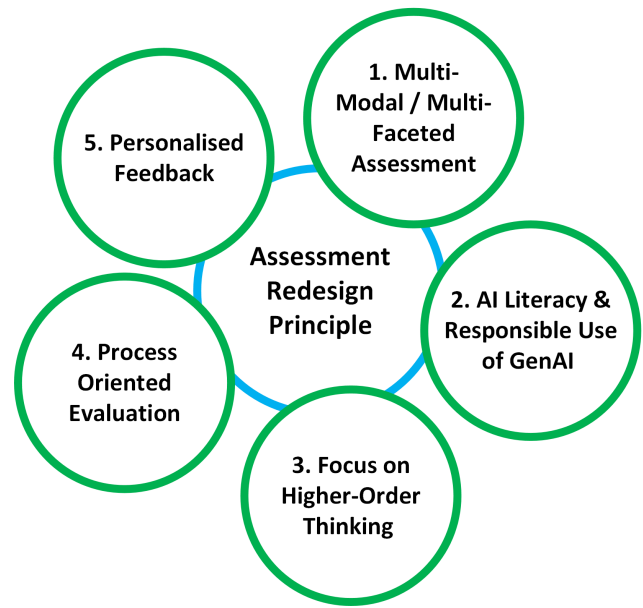

Fig. 1. Assessment redesign principles for Project-Based Assessment (PBA).

## IV. PRINCIPLES FOR REDESIGNING ASSESSMENT

The key concerns for PBA design in the era of GenAI are the product authenticity, academic integrity and learning validation [5]–[7], [25]. The redesign principles for PBA need to answer these questions, including the opportunity for students to use GenAI during the learning process. Figure 1 illustrates five principal elements for PBA redesign, which handle concerns raised by GenAI and also traditional assessment design requirements.

### A. Multi-Modal / Multi-Faceted Assessment

Multi-modal and multi-faceted assessment offers a robust approach to evaluating student learning by incorporating diverse evidence types. The evidence types may vary depending on the project types. Some examples are written reports, presentations, code artifacts, and reflective statements. This method enhances product authenticity by aligning assessments with real-world tasks and professional outputs, while also supporting academic integrity by making it more difficult to outsource or fabricate work. By triangulating assessment through multiple formats and perspectives, educators can better validate the originality, depth, and individual contribution of student efforts in the age of GenAI.

To incorporate the multi-modal and multi-faceted assessments, the following key points can be considered in the PBA design:

- Multi-modal assessment combines portfolios, journals, presentations, posters, peer and self-assessment, reports and progressive supervisor evaluation,
- Multi-faceted approach includes student learning evaluation from various viewpoints. The evaluation may come from various stakeholders, including project clients, academic supervisor, and industry supervisor.

The assessment should consider a variety of proposed or implemented approaches and solutions produced for the project work, and
- Assessments may consider process documentation, oral defences, and in-class components/activities. Ensure that assessments are prepared for continuous evaluation and include checkpoints throughout the learning process.

### B. AI Literacy & Responsible Use of GenAI

Assessing students' understanding of AI literacy and responsible use of GenAI is the key for future graduates to be fit for the industry and society. As the PBA are usually at the end of the program, assessing students' ability in this area is critical for PBA. To incorporate AI literacy and responsible use of GenAI, the following key points can be considered in the PBA design:
- Documenting the students' journey of reflection on the use of GenAI in PBA,
- Setting expectations about usage of GenAI according to the relevant institute's policies and the project's guidelines, and
- Reflect on understanding of ethical use of GenAI and AI-bias awareness.

### C. Focus on Higher-Order Thinking

The lower-order activities and tasks can be completed with the aid of the GenAI tools with limited student effort. Therefore, incorporating the assessment tasks with higher-order thinking in PBA is crucial in the era of GenAI for learning validation. The assessment should be designed in such a way that it makes sure that students' ability to "think with GenAI" rather than "letting GenAI think for you" is key. To incorporate higher-order thinking, the following key points can be considered in the PBA design:
- Checking the given assessment tasks or problems with the recent GenAI tools for possible completion with limited efforts. This process makes sure that students are required to think about the given tasks or problems and make efforts to solve the given tasks, even while using GenAI tools, and
- Assessments should judge students' creativity while solving the problems, critical thinking components, and teamwork collaboration.

### D. Process-Oriented Evaluation

Traditional assessment methods often emphasise the final product, and the product is now significantly influenced or impacted, or created by GenAI. Therefore, there is a need to shift assessment from an overwhelming focus on the final product to a more holistic view that values the learning journey, the decisions undertaken, the challenges navigated, and the skills acquired during the project lifecycle [26], [27]. This also needs to involve analysing student interactions with GenAI tools (where appropriate and with privacy considerations). Therefore, the focus of the assessment should be valuing and assessing the learning journey, iteration, decision-making processes, including how GenAI was used as a tool through multi-modal assessments, as discussed in previous sections.

To incorporate the process-oriented evaluation, the following key points can be considered in the PBA design:
- Multi-stage assessment to observe the progress either at the subject level or even at the program level,
- Documenting the Journey of student learning to ensure the authenticity of the product,
- Documenting the process, reflection, decision-making, and ethical use of GenAI,
- Ensure that the assessments are continuous with checkpoints, and
- Documenting and reporting on the analysis of student interactions with GenAI tools.

### E. Personalised Feedback

The observation of students' learning is crucial for learning validation in the era of GenAI. The evaluation of the use of GenAI-based tools or a supervisor for quick and personalised feedback on project-based activities during the learning journey, and the utilisation or uptake of that feedback in project progress, is key [28].
- Evaluate students' reflection on the feedback provided and how that feedback is addressed in the project, and
- Supervisor should observe and evaluate the use of provided feedback during the PBL process to ensure that students' progress on the learning.

Next, we will discuss the assessment model based on the assessment redesign principle.

## V. PROPOSED ASSESSMENT MODEL

The proposed assessment model based on the defined redesigned principles in Section IV, is illustrated in Figure 2. The conceptual model for PBA in the era of GenAI has four major components: *Assessment redesigned principles*; *Project Lifecycle & Assessment Hub*; *Evaluation Viewpoints*, and *Six Elements of PBA*.

The "Project Lifecycle & Assessment Hub" at the centre of the figure represents the student's journey through their PBL. The hub is informed by assessment redesign principles as discussed in Section IV and two evaluation viewpoints: traditional focus and GenAI insight. All six elements of PBA (shown on the left and right sides of the hub) are designed based on the assessment redesign principles and two evaluation viewpoints. In "GenAI Insight", we use assessment to evaluate the use of GenAI with how effectively, critically, and ethically the student engaged with it and what unique human skills (critical thinking, creativity, evaluation) they demonstrated in conjunction with, or independent of, the GenAI. These broader changes fundamentally alter how to approach PBA.

### A. Project Lifecycle & Assessment Hub

The "Project Lifecycle & Assessment Hub" at the centre of the figure illustrates the student's progression through their PBL experience. This hub is shaped by the principles of assessment redesign outlined in Section IV, and is viewed through two complementary lenses: a traditional assessment perspective and a GenAI-informed perspective.

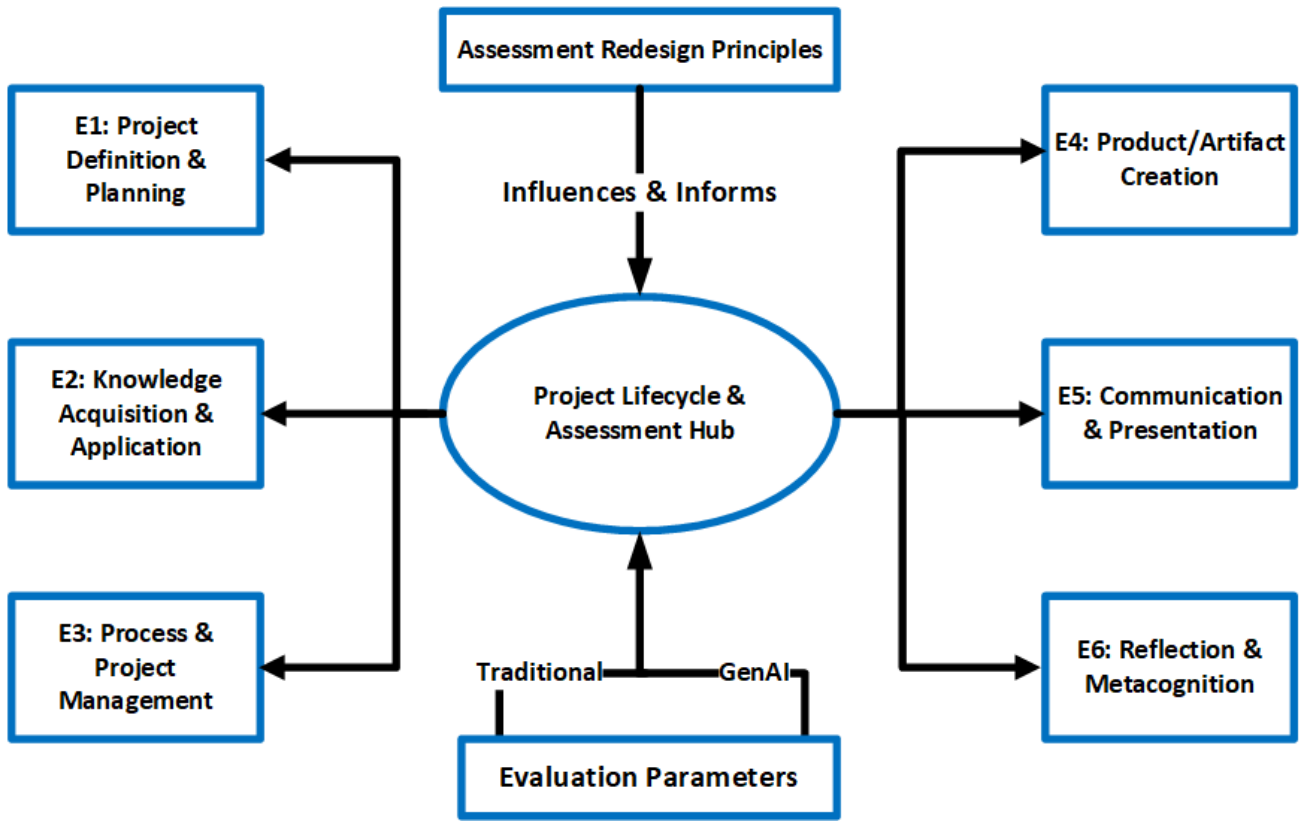

Fig. 2. Conceptual model for PBA in the era of GenAI.

### B. Evaluation: Traditional Focus and GenAI Insight

The PBA should be evaluated from two viewpoints: the Traditional perspective and GenAI insight. In the traditional view, the evaluation is focused on the assessment evaluation techniques of PBA as before the GenAI era, with some minor modifications, and the modifications will be highlighted in the next section.

In "GenAI Insight", we use assessment to evaluate the use of GenAI with how effectively, critically, and ethically the student engaged with it and what unique human skills (critical thinking, creativity, evaluation) they demonstrated in conjunction with, or independent of, the GenAI. These broader changes fundamentally alter how to approach assessment design in the entire PBL process. Next, we will discuss the six elements of PBA, considering assessment redesign principles and both evaluation viewpoints.

### C. Project-Based Assessment (PBA) Elements and Evaluation

In this section, the six PBA elements (E1-E6), along with two evaluation viewpoints, are discussed.

*1) Project Definition and Planning (E1) – With GenAI as a Potential Aid:* In this element, students can use GenAI as a potential aid for brainstorming, idea refinement, and resource discovery. Traditionally, the assessment should evaluate the clarity of the driving question, problem statements, feasibility of the project plan, identification of learning goals, and initial research in this element. The initial research may be potentially assisted by GenAI.

In GenAI Consideration, the assessment should evaluate the student's ability to formulate effective prompts for GenAI in idea generation/brainstorming, scope or idea refinement, and resource discovery, and their ability to judge or critically evaluate in selecting and adapting GenAI suggestions according to the project' needs and clarity on the transparency in GenAI usage for the project work.

*2) Knowledge Acquisition and Application (E2)– Potentially Leveraging GenAI:* In this element, students can potentially leverage GenAI for knowledge acquisition. Traditionally, the assessment is focused on assessing the depth of understanding of core concepts, research skills (using diverse sources, including traditional sources and GenAI-generated information), and the ability to apply knowledge gained from research to solve the project's central problem in this element.

In GenAI Consideration, the assessment should assess the student's capability to critically evaluate GenAI outputs for accuracy, bias, and completeness. Also, the assessment needs to evaluate the ability to synthesise GenAI-generated information with other sources effectively and also on the transparency of the use of GenAI.

TABLE I
PROJECT-BASED ASSESSMENT (PBA) ELEMENTS AND TWO EVALUATION VIEWPOINTS

| Element | Traditional Focus | GenAI Insight |
|---|---|---|
| Project Definition and Planning | <ul><li>Clarity of driving question</li><li>Feasibility of the project plan</li><li>Scope of the project</li><li>Identification of learning goals</li><li>Initial research</li></ul> | <ul><li>Skill in using GenAI for brainstorming or idea refinement</li><li>Effective prompt formulation for project activities</li><li>Critical evaluation of GenAI outputs</li><li>Transparency in the use of GenAI in the project</li></ul> |
| Knowledge Acquisition and Application | <ul><li>Depth of conceptual understanding</li><li>Research skills and using diverse sources</li><li>Application of knowledge</li><li>Ethical sourcing</li></ul> | <ul><li>Critical evaluation of GenAI-generated information for accuracy and bias</li><li>Synthesis of GenAI information with other ethical sources</li><li>Demonstrating research beyond GenAI</li><li>Acknowledgement on the use of GenAI sources</li></ul> |
| Process and Project Management | <ul><li>Team collaboration skills</li><li>Time management skills</li><li>Problem-solving and adaptability</li></ul> | <ul><li>Use of GenAI as an organisational/productivity tool</li><li>Student ownership of the process and not over-reliance on GenAI</li><li>Documentation of the process of GenAI interaction</li></ul> |
| Product/Artifact Creation | <ul><li>Quality of the final product</li><li>Innovation, creativity, fulfilment of requirements</li><li>Technical skills</li></ul> | <ul><li>Assessing students' unique contribution versus GenAI-generated portions</li><li>Student's role in guiding, curating, and significantly refining GenAI outputs</li><li>Originality and intellectual ownership</li><li>Appropriate acknowledgment of GenAI</li><li>Focus on aspects of deep context and novel synthesis where GenAI struggles</li></ul> |
| Communication and Presentation | <ul><li>Clarity, coherence, effectiveness in conveying outcomes</li><li>Learning, and audience engagement</li></ul> | <ul><li>Articulation of unique insights beyond GenAI</li><li>Authenticity of voice</li><li>Genuine comprehension demonstrated through question and answer</li></ul> |
| Reflection and Metacognition | <ul><li>Depth of self-assessment on learning</li><li>Identification of challenges/growth</li><li>Connections to broader concepts</li></ul> | <ul><li>Reflection on the role, benefits, and limitations of GenAI in the project</li><li>Understanding ethical implications of GenAI use</li><li>Future learning plans considering GenAI</li></ul> |

*3) Process and Project Management (E3) – GenAI as a Potential Organisational Tool:* In this element, students can use GenAI tools as a potential organisational tool. The assessment is focused on assessing collaboration and teamwork (assuming that the project is undertaken as a group work), time management, iteration and refinement of ideas, problem-solving during the project lifecycle, and adaptability in this element.

In GenAI Consideration, the assessment is focused on students' ability to leverage GenAI tools for project organisation, task management, or even generating initial drafts for review and significant revision. The focus of the assessment should remain on student ownership of the process (not over-reliance on GenAI) and documentation of process or progress, including GenAI interaction.

*4) Product/Artifact Creation (E4) – GenAI as a Potential Co-creation Tool:* In this element, students can use GenAI as a potential co-creation tool for product development. Traditionally, the assessment should be focused on assessing the quality of the final product/solution, technical skills earned, innovation, creativity, and fulfilment of project requirements.

In GenAI Consideration, GenAI is used in product creation (e.g., drafting text, generating code snippets, creating images and so on). The assessment should be focused on shifts to focus on the student's role in guiding the GenAI, curating and refining its output, and adding unique human value to the product creation. The assessment should focus on the originality and authenticity of the student's contribution to the project work. Evaluating the student's ability to appropriately cite or acknowledge GenAI use becomes crucial in the assessment design.

*5) Communication and Presentation (E5) – GenAI as a Potential Aid for Structuring/Refining:* In this element, students can use GenAI as a potential aid for structuring/refining the content for presentation and report writing. The focus of the traditional assessment for this element is to get clarity and effectiveness of communication (written, oral, visual), ability to articulate the learning journey and project outcomes, and engaging the audience.

In GenAI Consideration, assessment should focus on the coherence of their arguments, the depth of their understanding, and their ability to present their work effectively, and genuine comprehension demonstrated in question and answer defence.

*6) Reflection and Metacognition (E6) – Crucial in the GenAI Context:* This component is a vital element in the era of GenAI. In this element, we are assessing: Students' ability to critically reflect on their learning process; identify challenges and how they were overcome; and able to articulate future learning. Additionally, we need to evaluate the use of

GenAI with its benefits and limitations in the project context.

In GenAI insight, the proposed assessment should probe students' understanding of when and how to use GenAI ethically and effectively, their awareness of potential pitfalls (e.g., over-reliance, misinformation, hallucinated content, and epistemic unreliability), and their ability to learn from interacting with GenAI during the process. Table I illustrates a summary of six PBA elements with traditional and GenAI evaluation viewpoints.

TABLE II
A SAMPLE PBA DESIGN FOR A CAPSTONE SUBJECT

| Assessment Name | Weight (%) | Type | Due (Week) |
|---|---|---|---|
| Project Proposal | 10 | Group | 3 |
| Research & Planning Log – Formative | 5 and Hurdle | Group | Weekly |
| Peer Feedback / Collaboration Log | 10 | Individual | 5 & 12 |
| Interim Report | 10 | Group | 5 |
| Supervisor (Client and Academic) Evaluation | 10 | Individual | Weekly |
| Reflection Report on GenAI Use | 10 | Individual | 5 & 12 |
| Reflection Report on Teamwork or Personal Growth | 10 | Individual | 5 & 12 |
| Presentation With Viva | 10 | Individual | 6 and 11 |
| Final Report | 25 | Group | 12 |

## VI. USE CASE EXAMPLE AND ANALYSIS

In this section, we introduce a sample assessment design for an undergraduate capstone project subject, which is followed by its analysis in various aspects.

### A. Use Case Example

This section presents a sample assessment design for an undergraduate capstone project subject delivered over 12 weeks. The project is provided by an external client and completed collaboratively by a group of students. A sample set of assessments for a capstone subject based on the proposed GenAI-Era based Principles is illustrated in Table II. The Table shows the weight, type, and due date for assessment. The Research & Planning Log – Formative has a weight of 5%, but the student needs to pass this assessment to pass the subject. The strengths of the sample assessments are:

- Assessments like the project proposal, interim report, and final report simulate professional practices and encourage real-world problem-solving.
- The timeline distributes tasks across the semester (weeks three to 12), supporting progressive development and continuous engagement and continuous delivery. The use of logs and interim reports builds toward the final product.
- The inclusion of both a GenAI-specific reflection report and a general reflection report promotes critical thinking and ethical awareness of GenAI use in academic and professional settings.
- The design avoids common pitfalls of group work by incorporating: Individual reflection, peer feedback logs, supervisor evaluations, and individual presentations/viva.
- Weekly supervisor evaluations bring external perspectives into the assessment, offering formative feedback loops and increasing accountability.

### B. Analysis

The proposed set of assessments is analysed in terms of the proposed GenAI-Era Principles discussed in Section IV, TEQSA's assessment design principles in the era of GenAI discussed in Section I, and the PBA model introduced in Section V. Table III illustrates the mapping between GenAI-Era principles and evidence in the proposed sample assessment design for a capstone project subject. The mapping clearly shows how the assessment tasks are mapped to the principles. Therefore, each PBA design for the subject should complete the mapping of the assessment tasks with the proposed assessment principles.

This assessment design aligns with the Higher Education Standards Framework (Threshold Standards) 2021 set by the TEQSA [29] and assessment design guide in the era of GenAI [1], and also reinforces assessment integrity and security, particularly by:

- Diversified assessment methods (used in the sample PBA) enhance academic integrity and assessment security: A variety of assessment types (e.g., proposals, logs, peer feedback, supervisor evaluations, reflections, presentations, and reports). This multi-modal assessment strategy makes it harder for academic misconduct to occur consistently across different formats, thus enhancing assessment security and integrity.
- Formative logs and peer feedback (used in the sample PBA) promote authentic student engagement and enhance academic integrity: The inclusion of formative Research & Planning Logs and Peer Feedback/Collaboration Logs ensures continuous monitoring of student participation and learning processes. These iterative checkpoints reduce opportunities for last-minute fraudulent work and encourage authentic, traceable student contributions, thereby supporting academic integrity.
- Supervisor evaluations and viva (used in the sample PBA) add verification layers for assessment security: The supervisor (client and academic) evaluation and presentation, along with the viva, act as strong integrity checkpoints by involving external authenticators (supervisors) and oral defence mechanisms (viva). These methods verify the student's ownership of their work and serve as a deterrent against impersonation or outsourced submissions, thereby reinforcing assessment security. Learning resources and educational support are provided through the use of supervisor feedback and formative tasks.
- Address risk associated with GenAI-related academic integrity by explicit engagement with GenAI via reflective practice.

TABLE III
ANALYSIS ON PBA DESIGN ALIGNMENT WITH GENAI-ERA PRINCIPLES

| Assessment Principle Elements | Evidence in Capstone Assessment Design |
| --- | --- |
| Multi-Modal and Multi-Faceted Assessments | There is a diversity of formats: written reports (*Proposal, Interim, Final*), reflective writing (*Reflection reports*), oral components (*Presentation with viva*), and peer/supervisor inputs. This provides a holistic view of student capabilities. |
| AI Literacy and Responsible Use of GenAI | The dedicated *Reflection report on GenAI use* ensures that students engage critically with generative AI, addressing ethics, transparency, and capability boundaries. It encourages mindful, responsible integration of GenAI tools in academic work. |
| Focus on Higher-Order Thinking | Tasks like the *Project proposal*, *Final report*, and *Presentation with viva* require problem-solving, synthesis, and original contribution — moving beyond knowledge recall to application, analysis, and evaluation. |
| Process-Oriented Evaluation | The assessment includes continuous elements such as the *Research & planning log (weekly)*, *Supervisor evaluation*, and two-stage *Reflection reports*. These support tracking student development over time, not just the final output. |
| Personalised Feedback | The *Supervisor evaluation* (weekly) and peer feedback cycles enable continuous, individualised input. *Viva presentations* also allow for personal verification and feedback in a dialogic format. |

TABLE IV
ANALYSIS ON PBA DESIGN ALIGNMENT WITH PROJECT ELEMENTS AND EVALUATION PARAMETERS

| Assessment Name | Evaluation Parameter(s) | PBA Element(s) |
| --- | --- | --- |
| Project Proposal | Traditional Focus | Project Definition and Planning (E1) |
| Research & Planning Log | Both | Process and Project Management (E3), Product/Artifact Creation (E4) |
| Peer Feedback / Collaboration Log | Both | Process and Project Management (E3), Communication and Presentation (E5) |
| Interim Report | Traditional Focus | Project Definition and Planning (E1), Knowledge Acquisition and Application (E2), Process and Project Management (E3), Product/Artifact Creation (E4), Communication and Presentation (E5) |
| Supervisor (Client and Academic) Evaluation | Both | Project Definition and Planning (E1), Knowledge Acquisition and Application (E2), Process and Project Management (E3), Product/Artifact Creation (E4), Communication and Presentation (E5), Reflection and Metacognition (E6) |
| Reflection Report on GenAI Use | GenAI Insight | Project Definition and Planning (E1), Knowledge Acquisition and Application (E2), Process and Project Management (E3), Product/Artifact Creation (E4), Communication and Presentation (E5), Reflection and Metacognition (E6) |
| Reflection Report on Teamwork or Personal Growth | Traditional Focus | Knowledge Acquisition and Application (E2), Process and Project Management (E3), Product/Artifact Creation (E4), Communication and Presentation (E5), Reflection and Metacognition (E6) |
| Presentation With Viva | Traditional Focus | Communication and Presentation (E5) |
| Final Report | Traditional Focus | Project Definition and Planning (E1), Knowledge Acquisition and Application (E2), Process and Project Management (E3), Product/Artifact Creation (E4), Communication and Presentation (E5) |

The sample PBA design introduced is also mapped with project elements and evaluation parameters. The mapping analysis covers how various assessments use evaluation parameters and covers PBA elements introduced in Section V. This analysis is presented in Table IV. It highlights the integration of traditional and GenAI Insight evaluation approaches introduced across different project elements, ensuring comprehensive assessment coverage. Additionally, the table shows how the sample assessments are mapped to PBA elements.

Therefore, the proposed assessment design addresses emerging risks associated with GenAI while also meeting traditional assessment practices, including assessment security and integrity set by the regulatory agency.

## VII. CONCLUSIONS AND FUTURE DIRECTIONS

In conclusion, as GenAI continues to reshape educational practices, assessment models must evolve to uphold academic integrity while fostering learning. By shifting the focus toward process-oriented evaluation, Scaffolding of assessments, critical engagement, reflection and personalised feedback, the proposed model offers a forward-looking approach that aligns with the realities of GenAI-enhanced learning environments. This ensures that assessments not only remain valid and authentic but also support the development of essential skills for future-ready graduates. Therefore, the academics recommend using the proposed model for PBA to ensure student learning and assessment security.

As future work, we plan to implement this PBA model within the capstone project subject and systematically evaluate both student and instructor responses to its effectiveness, usability, and impact on learning outcomes.

## REFERENCES

[1] S. H. Jason M Lodge and M. Bearman, "Assessment reform for the age of artificial intelligence," https://www.teqsa.gov.au/sites/default/files/2023-09/assessment-reform-age-artificial-intelligence-discussion-paper.pdf, 2023, accessed: (03-06-2025).

[2] Q. Xia, X. Weng, F. Ouyang, T. J. Lin, and T. K. Chiu, "A scoping review on how generative artificial intelligence transforms assessment in higher education," *International Journal of Educational Technology in Higher Education*, vol. 21, no. 1, p. 40, 2024.

[3] N. Boughattas, W. Neji, and F. Ziadi, "Project based assessment in the era of generative ai-challenges and opportunities," *Tsvetkova, Anastasia; Morariu, Andrei-Raoul; Hellström, Magnus; Bolbot, Victor; Virtanen, Seppo Investigation of student perspectives on curriculum needs for autonomous shipping*, p. 347, 2024.

[4] C. Combrinck and N. Loubser, "Student self-reflection as a tool for managing genai use in large class assessment," *Discover Education*, vol. 4, no. 1, p. 72, 2025.

[5] B. Divjak, B. Svetec, and K. Pažur Aničić, "PBL Meets AI: Innovating Assessment in Higher Education," in *CSEDU 2025 17th International Conference on Computer Supported Education Proceedings (Volume 2)*. Setúbal: SCITEPRESS, 2025, pp. 120–130.

[6] D. Smith and N. Francis, "Process not product in the written assessment," in *Using generative AI effectively in higher education*. Routledge, 2024, pp. 115–126.

[7] C. Gonsalves, "Contextual Assessment Design in the Age of Generative AI." *Journal of Learning Development in Higher Education*, 2025.

[8] M. Belkina, S. Daniel, S. Nikolic, R. Haque, S. Lyden, P. Neal, S. Grundy, and G. M. Hassan, "Implementing generative AI (GenAI) in higher education: A systematic review of case studies," *Computers and Education: Artificial Intelligence*, p. 100407, 2025.

[9] R. Kadel, B. K. Mishra, S. Shailendra, S. Abid, M. Rani, and S. P. Mahato, "Crafting tomorrow's evaluations: assessment design strategies in the era of generative AI," in *2024 International Symposium on Educational Technology (ISET)*. IEEE, 2024, pp. 13–17.

[10] A. Sharma, S. Shailendra, and R. Kadel, "Experiences with Content Development and Assessment Design in the Era of GenAI," in *2025 6th International Conference on Computer Science, Engineering, and Education (CSEE)*, 2025, pp. 1–5.

[11] EDUCAUSE, "Higher Education Generative AI Readiness Assessment," Apr. 2024, accessed: 2025-06-02. [Online]. Available: https://library.educause.edu/resources/2024/4/higher-education-generative-ai-readiness-assessment

[12] S. Shailendra, R. Kadel, and A. Sharma, "Framework for adoption of generative artificial intelligence (GenAI) in education," *IEEE Transactions on Education*, 2024.

[13] R. Jongkind, E. Elings, E. Joukes, T. Broens, H. Leopold, F. Wiesman, and J. Meinema, "Is your curriculum GenAI-proof? A method for GenAI impact assessment and a case study," *MedEdPublish*, vol. 15, no. 11, p. 11, 2025.

[14] M. Hornberger, A. Bewersdorff, D. S. Schiff, and C. Nerdel, "A multinational assessment of AI literacy among university students in Germany, the UK, and the US," *Computers in Human Behavior: Artificial Humans*, vol. 4, p. 100132, 2025.

[15] X. Gu and B. J. Ericson, "AI Literacy in K-12 and Higher Education in the Wake of Generative AI: An Integrative Review," *arXiv preprint arXiv:2503.00079*, 2025.

[16] K. A. LaFlamme, "Scaffolding AI literacy: An instructional model for academic librarianship," *The Journal of Academic Librarianship*, vol. 51, no. 3, p. 103041, 2025.

[17] K. Sol, S. Sok, and K. Heng, *Rethinking Assessment in Higher Education in the Age of Generative AI*. Singapore: Springer Nature Singapore, 2025, pp. 1–5. [Online]. Available: https://doi.org/10.1007/978-981-13-2262-4_327-1

[18] H. Desai, "What's worth measuring? The future of assessment in the AI age," *UNESCO*, May 2025, accessed: 2025-06-02. [Online]. Available: https://www.unesco.org/en/articles/whats-worth-measuring-future-assessment-ai-age

[19] M. Kassorla, M. Georgieva, and A. Papini, "Defining AI Literacy for Higher Education," https://www.educause.edu/content/2024/ai-literacy-in-teaching-and-learning/defining-ai-literacy-for-higher-education, October 2024, accessed: 2025-06-02.

[20] Digital Education Council, "What Faculty Want: Key Results from the Global AI Faculty Survey 2025," Jan. 2025, accessed: 2025-06-02. [Online]. Available: https://www.digitaleducationcouncil.com/post/what-faculty-want-key-results-from-the-global-ai-faculty-survey-2025

[21] Australian Council for Educational Research. (2025, January) Ai in education: ensuring teacher agency in a technology-empowered world. Accessed: 2025-06-05. [Online]. Available: https://www.acer.org/au/discover/article/ai-in-education-ensuring-teacher-agency-in-a-technology-empowered-world

[22] E. du Plessis, "Embracing project-based assessments in the age of ai in open distance e-learning," *International Journal of Information and Education Technology*, vol. 15, no. 2, pp. 372–381, February 2025. [Online]. Available: https://www.ijiet.org/vol15/IJIET-V15N2-2249.pdf

[23] R. Williams, S. Ali, N. Devasia, D. DiPaola, J. Hong, S. P. Kaputsos, B. Jordan, and C. Breazeal, "Ai+ ethics curricula for middle school youth: Lessons learned from three project-based curricula," *International Journal of Artificial Intelligence in Education*, vol. 33, no. 2, pp. 325–383, 2023.

[24] P. Mueller-Csernetzky, E. Malakhatka, L. Thuvander, D. Kragic, and J. Kabo, "Technological frontiers in education: Exploring the impact of ai and immersive learning," in *Human-Technology Interaction: Interdisciplinary Approaches and Perspectives*. Springer, 2025, pp. 291–328.

[25] M. S. Torkestani and T. Mansouri, "Machine vs machine: Using ai to tackle generative ai threats in assessment," in *Proceedings of The Chartered Association of Business Schools (CABS)*, 2025, author accepted manuscript, University of Exeter Institutional Repository, handle 10871/141147. [Online]. Available: https://ore.exeter.ac.uk/repository/handle/10871/141147

[26] C. B. Hodges and P. A. Kirschner, "Innovation of instructional design and assessment in the age of generative artificial intelligence," *TechTrends*, vol. 68, no. 1, pp. 195–199, 2024.

[27] D. B. Guruge and R. Kadel, "Towards an Holistic Framework to Mitigate and Detect Contract Cheating within an Academic Institute—A Proposal," *Education Sciences*, vol. 13, no. 2, p. 148, 2023.

[28] S. Rutherford, C. Pritchard, and N. Francis, "Assessment is learning: developing a student-centred approach for assessment in higher education," *FEBS Open Bio*, vol. 15, no. 1, pp. 21–34, 2025.

[29] Australian Government, "Higher Education Standards Framework (Threshold Standards) 2021," Apr. 2021, registered on 27 April 2021. Accessed: 2025-06-02. [Online]. Available: https://www.legislation.gov.au/F2021L00488/latest/text